\documentclass[twocolumn,preprintnumbers,amsmath,aps]{revtex4-1}

\usepackage{graphicx}
\usepackage{dcolumn}
\usepackage{bm}
\usepackage{xcolor}

\begin{document}






\title{Quantum smell: tunneling mechanisms in olfaction}






\author{Dominik Szcz{\c{e}}{\'s}niak$^{1}$}
\email{d.szczesniak@ujd.edu.pl}
\author{Ewa A. Drzazga-Szcz{\c{e}}{\'s}niak$^{2}$}
\author{Sabre Kais$^{3, 4}$}


\affiliation{
${^1}$Institute of Physics, Faculty of Science and Technology, Jan D{\l}ugosz University in Cz{\c{e}}stochowa, 13/15 Armii Krajowej Ave., 42200 Cz{\c{e}}stochowa, Poland \\
${^2}$Department of Physics, Faculty of Production Engineering and Materials Technology, Cz{\c{e}}stochowa University of Technology, 19 Armii Krajowej Ave., 42200 Cz{\c{e}}stochowa, Poland \\
${^3}$Department of Electrical and Computer Engineering, North Carolina State University, Raliegh, North Carolina 27606, United States\\
${^4}$ Department of Chemistry, North Carolina State University, Raleigh, NC 27695}

\date{\today}


\begin{abstract}

The mechanism by which odorants are recognized by olfactory receptors remains primarily unresolved. While charge transport is believed to play a significant role, its precise nature is still unclear. Here, we present a novel perspective by exploring the interplay between the intrinsic energy scales of odorant molecules and the gap states that facilitate intermolecular charge transport. We find that odorants act as weak tunneling conductors mainly because of the limited magnitude of electronic coupling. This behavior is further connected to electron-phonon interactions and reorganization energy, suggesting that physically meaningful values for these parameters emerge only in the deep off-resonant tunneling regime. These findings complement the swipe card model of olfaction, in which an odorant needs both the right shape to bind to a receptor and the correct vibrational frequency to trigger signal transduction. Moreover, they reveal that the underlying mechanisms are much more complex than previously assumed.



\end{abstract}

\maketitle







As far as human perception is concerned, the sense of smell, or olfaction, is one of the first concepts that comes to our minds. However, the obviousness of its existence does not go hand in hand with a complete understanding of why we are able to perceive a plethora of various scents. Although it is already well known that it takes activation of the odorant receptors to stimulate neurons in our noses \cite{buck1991}, the underlying mechanism of olfactory discrimination is somewhat unknown. So far, several factors have been identified as potentially pivotal for our smell impressions, namely the physicochemical attributes (size, shape, bonding etc.) \cite{moncrieff1949, amoore1963, billesbolle2023} and vibrations \cite{dyson1928, wright1954, turin1996} of the odorant molecules. The debate on which of these features is dominant, or even relevant, in the olfactory process appears to be at the cornerstone of the future theory of smell \cite{vosshall2015, hoehn2018}. Unfortunately, none of them seems to be sufficiently explored, including their likely interdependencies.

All the above intricacies of smell naturally lead to an essential question: can we arrive at a unified picture of the olfaction mechanism that allows for simultaneous experimental verification? Herein, it is argued that an important insight in this respect can be obtained by recalling the general interpretation of charge transport at the quantum level. As may already be apparent, such reasoning shares some similarities with the vibrational theory of smell \cite{turin1996}. Still, it does not pre-assume the validity of this idea, being rather aimed at assessing the general plausibility and role of electron transport in shaping olfaction. In fact, this process is cardinal to the neural system and as such naturally calls for inclusion in the analysis of the olfactory segments. This is to say, even if electron transport is unable to solely explain the olfaction mechanism, it can still be expected to be of fundamental importance to this problem.

In detail, when considering charge transport in the quantum regime, we intuitively reflect on the phenomenon of electrons scattering from a potential barrier. This effect lays the foundations for our understanding of electronic transport in mesoscopic systems, as formulated by Landauer \cite{landauer1957}. While the aforementioned approach originally relies on electrons being propagated in the semi-infinite terminals before reaching a scatterer, it seems to be viable for fully molecular considerations as well. By following in spirit the studies of Nitzan {\it et al.} \cite{davis1997, segal2000, nitzan2001, nitzan2002}, it can be deduced that the electronic transport across a molecule coupled to a periodic medium mimics, to a great extent, the tunnelling process in the donor-molecule-acceptor system. Hence, the schematic representation of an odorant docked in the olfactory receptor. This point of view is reinforced by the fact that in the zero bias limit ({\it i.e.} in the regime where Nitzan's argumentation appears to be particularly valid \cite{nitzan2001}), electron tunnelling is largely independent of the considered terminals \cite{samanta1996}. To be specific, this is the non-resonant tunnelling regime, where electron transport is proportional to $e^{-{\beta}L}$, with $\beta$ being the decay rate and $L$ denoting the scatterer length. Although the above derives from partially periodic considerations, it is in qualitative agreement with what can be found for the electron exchange rates between purely molecular species \cite{marcus1956, mcconnell1961, valianti2019}.

The presented rationale means that electron tunnelling in the odorant-receptor system may be conveniently captured within the Landauer approach. According to the mentioned decay characteristics of this process, of pivotal importance to such discussion seem to be the exponent observables. The scatterer length, being the first of them, can be simply interpreted as the distance between the acceptor and donor sites in the olfactory receptor. This parameter is naturally limited by the ability of electrons to traverse the region between the receptor terminals in the tunnelling regime. The second feature is much more complex, as it captures the intrinsic properties of the bridging structure such as its chemical structure and energetics. Still, contrary to the available semi-classical \cite{marcus1956, brookes2007} or even quantum-mechanical \cite{solovyov2012, hoehn2015, liza2019, zhang2022} studies of the olfaction problem, the provided reasoning appears to have a more canonical character. This is not only due to the fact that it stems from the fundamental depiction of electron transport at the quantum level but also attempts to cut out secondary aspects such as the structure of acceptor and donor sites.

In this context, it is suggested that one of the two primary scenarios may occur, namely intra- or extra-molecular transport, both of which were initially considered by Brookes \cite{brookes2007, brookes2012, marcus1956}. Out of these two, the second approach was quickly disregarded due to the large energy gap of odorants, without any further analysis conducted \cite{brookes2012}. However, sizable band gaps in molecular systems do not necessarily lead to negligible tunnelling rates, as has been shown {\it e.g.} for alkanes \cite{tomfohr2002}. On the other hand, the analysis of tunnelling processes across DNA sequences suggests that, although their band gaps are not exceptionally large, the corresponding decay rates are high, leading to poor charge transport \cite{wang2004}. In what follows, it is argued here that the charge transport across odorant molecules should not be discussed solely based on the energy gap scale, without taking into account other important factors. Of particular importance is the coupling between electronic states and their interaction with molecular vibrations, which cannot be ruled out. Specifically, the former facilitates tunnelling and defines the strength of the corresponding transport channel, while the latter can be expected to further assist in mediating charge transport between donor and acceptor sites. That is to say, the vibrations of the odorant molecule do not merely have to bridge the energy gap, but can instead couple to the electrons responsible for shaping tunnelling across the associated potential barrier. Notably, this insight also constitutes a viable platform for experimental verification of the tunnelling mechanism due to the similarities between purely molecular systems and those terminated by metal electrodes, as discussed by Nitzan \cite{salomon2003, davis1997, segal2000, nitzan2001, nitzan2002}.

\begin{figure}[ht!]
\includegraphics[width=\columnwidth]{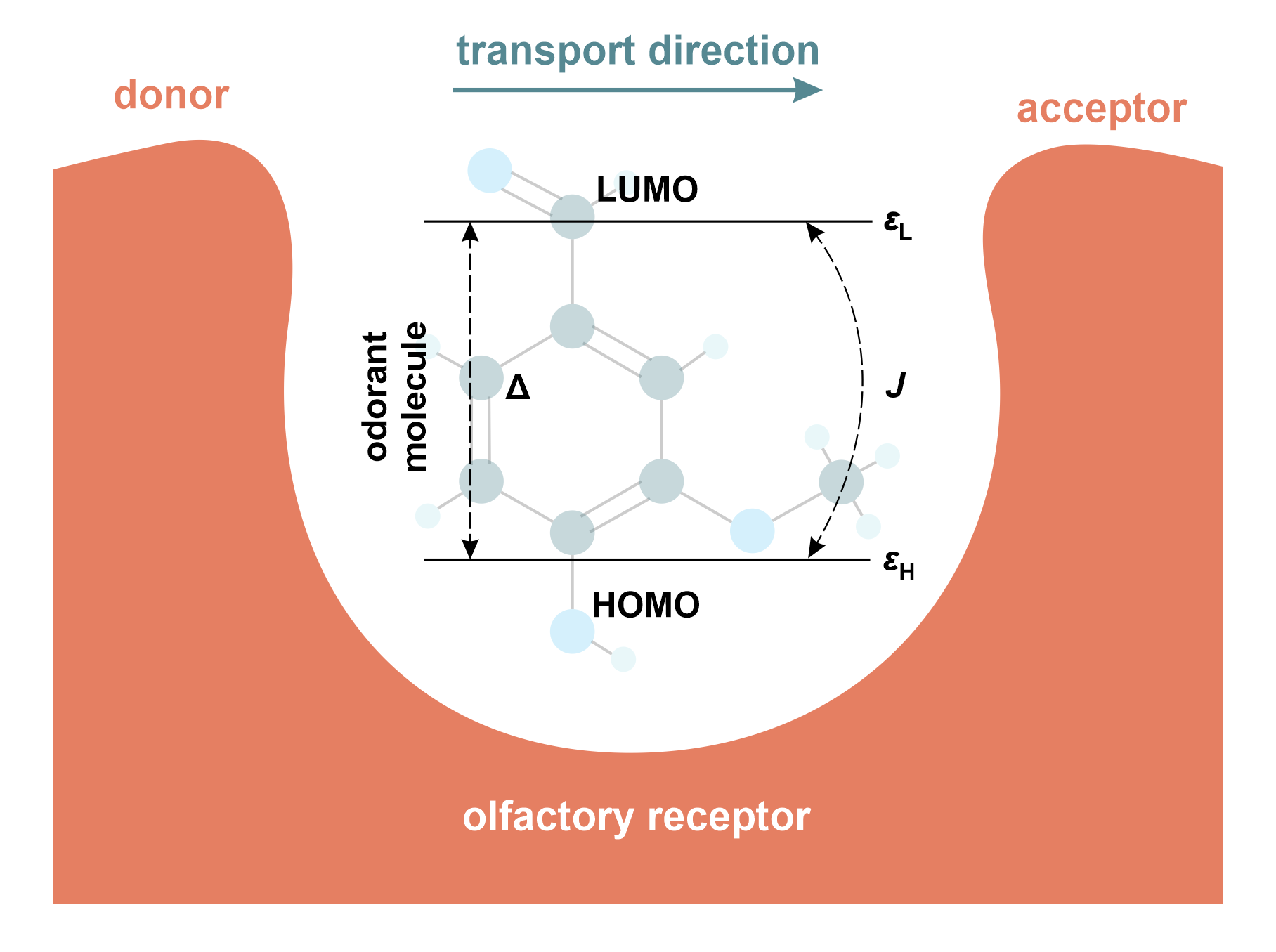}
\caption{Schematic representation of an odorant molecule locked in a olfactory receptor. The former is presented here as a two-level molecule defined by the highest-occupied (HOMO) and lowest-unoccupied (LUMO) molecular orbitals with energies $\varepsilon_{H}$ and $\varepsilon_{L}$, respectively.}
\label{fig01}
\end{figure}

The above can be pictured by considering interplay between mentioned energy scales. These are conveniently captured by the following Hamiltonian which yields the highest-occupied (HOMO) and lowest-unoccupied (LUMO) molecular orbitals of an odorant:
\begin{equation}
\label{eq1}
H= \varepsilon_{H} c^{\dagger}_{H} c_{H} + \varepsilon_{L} c^{\dagger}_{L} c_{L} + J \left( c^{\dagger}_{H} c_{L} + c^{\dagger}_{L} c_{H} \right),
\end{equation}
where $c_{H/L}$ $(c^{\dagger}_{H/L})$ creates (anihilates) electron with energy $\varepsilon_{H/L}$ at the HOMO/LUMO level and $J$ denotes electronic coupling between considered energy levels. In this framework, the energy gap is simply given as $\Delta=\epsilon_{L}-\epsilon_{H}$. The schematic representation of the odorant molecule is presented in Fig. \ref{fig01}.

With the above in mind, the tunnelling across an odorant molecule may be viewed as an evanescent wave coupling effect \cite{reuter2017, szczesniak2020}. As such, the evanescent modes mimic channels that enable the single-step transport of electrons across the molecule. Importantly, these can be effectively probed in the experiment by using {\it e.g.} the scanning tunnelling microscopy \cite{langlais1999}. In practice, there are several ways to characterize such evanescent waves, where one of them is the complex band structure technique \cite{tomfohr2002, szczesniak2016, reuter2017}. In this framework, the molecular region is considered to be infinitely extended in order to recover the decay characteristics of electrons. The central aspect of this analysis is the branch point of gap states at which $d\beta/dE=0$ and the molecule remains charge neutral. At this point, the decay rate corresponds to the direct tunnelling, in contrast to the indirect processes occurring at the remaining energies within the gap \cite{kane1961}.

\begin{figure}[ht!]
\includegraphics[width=\columnwidth]{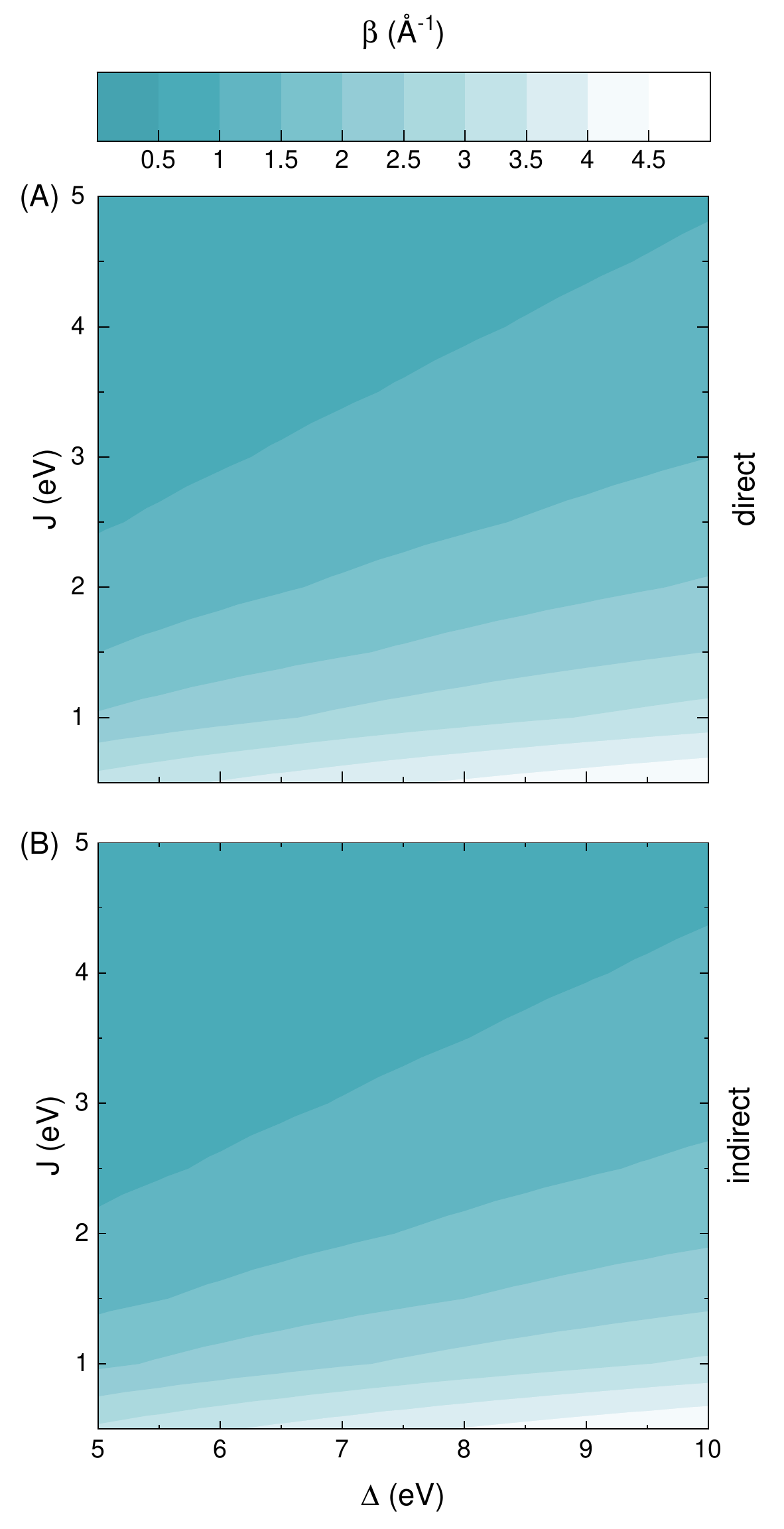}
\caption{The decay rates for odorant molecule as a function of electronic coupling and energy gap size. Two tunnelling regimes are considered, the direct (A) and indirect (B) one.}
\label{fig02}
\end{figure}

In Fig. \ref{fig02}, the magnitude of the decay rates for the direct and indirect tunnelling processes is depicted as a function of the electronic coupling and energy gap. The presented contour plots are derived from Eq. (\ref{eq1}) by using the complex band structure technique over a wide range of energy parameter values \cite{brookes2012, zhang2022, adams2022}. These are specifically chosen to span energy characteristics typical of odorant molecules, with only the coupling constant additionally extending toward higher values for a wider context. The energy of the HOMO level is set to zero and considered a reference value. In this manner, while the direct tunnelling rates correspond exactly to the branch point of gap states, the results for the indirect process are associated with an energy level equal to $\varepsilon_{L}/6$ {\it i.e.} close to the HOMO level. Note that, in the present analysis, the HOMO and LUMO are equidistant from the branch point and it is enough to sample for indirect tunnelling rates only at energies below the branch point.


A key observation from the results presented in Fig.~\ref{fig02} is that, within the assumed range of Hamiltonian parameters, the decay rates take values from 0.5~\AA$^{-1}$ or less up to nearly 5~\AA$^{-1}$. This wide variation indicates that the character of charge transport transitions from being weakly dependent on distance to exhibiting exponential decay with strong distance dependence. Such a shift has a significant impact on the underlying transport regime, which is indicative of a near-resonant process when decay rates are low (\(\beta < 0.5~\text{\AA}^{-1}\)) and suggests coherent and fully off-resonant tunnelling for higher decay values \cite{chen2021}. Interestingly, it can also be observed that decay rates exhibit relatively little sensitivity to changes in the energy gap under weak coupling conditions and yield low values only for strong electronic coupling. This is somewhat expected, since higher electronic coupling should promote more efficient charge transport, as suggested by the stronger overlap of orbitals. However, this behavior stands in partial contrast to the preliminary argument by Brookes, which suggests that tunnelling across an odorant molecule would be ineffective mainly due to the large energy gap characteristic of such molecules \cite{brookes2012}. In fact, the calculated decay rates at the midpoint value of electronic coupling become relatively efficient for transport to occur and only improve as the coupling increases. Moreover, these findings are mostly uniform, whether the direct or indirect process is considered. Only some small improvements in decay rate values are present in terms of the indirect tunnelling, but rather in the strong coupling limit.

In what follows, to estimate the regime of charge transport across odorants, it is essential to consider the interplay between the band gap and electronic coupling. Specifically, for the HOMO and LUMO to remain strongly localized, as in a typical odorant molecule, the corresponding electronic coupling must be significantly smaller than the band gap. According to Eq. (\ref{eq1}), it can be expected that hybridization between these two distinct states will be preserved when the coupling $J<1$ eV. Under such conditions, the results presented in Fig.~\ref{fig02} clearly suggest that charge transport via odorants likely falls into the deep off-resonant tunnelling regime. In other words, the simple exponential relation for the tunnelling breaks, the corresponding transmission becomes unmeasurably small, and other mechanisms (like hopping or phonon-assisted tunnelling) may dominate. However, this also indicates that odorant molecules are poor tunnelling conductors not only due to the large energy gaps but also to small electronic coupling. In fact, the presented results suggest that the electronic coupling plays a more significant role in suppressing tunnelling rates in odorants than the energy gaps themselves. Altogether, this implies that the extra-molecular mechanism proposed by Brookes \cite{brookes2012} remains viable, although the underlying rationale appears to be more complex than previously anticipated.

In this context, it is essential to discuss the relation between tunnelling and the vibrations of the odorant molecule. In the framework of Eq. (\ref{eq1}), the odorant can be regarded as a two-level vibronic system, or oscillating dimer, in which the HOMO and LUMO states are linked by a phonon-modulated tunnelling amplitude. In this first-order approximation, the effect of the vibrational mode is distilled into a single parameter of the electron–phonon interaction strength, as derived from the electronic coupling. Specifically, in the frozen-phonon regime, the linear electron-phonon coupling can be written as $\alpha=dJ/du$, with $u$ standing for the displacement per atom. To introduce the necessary dependence of electronic coupling on the distance, this parameter can be assumed to have the character of a two-central Slater-Koster integral that decays exponentially with the distance as $J(l)=J_{0}e^{-\eta(u/a_{0}-1)}$, where $J_{0}$ $(a_{0})$ is the equilibrium coupling (distance) between two sites and $\eta$ denotes the fall-off rate, which can be interpreted as a phonon mode softening/hardening rate \cite{naumis2017, szczesniak2025}. Note that the exponential form is one of the many available ones, but it is chosen here since it was proved to be well-suited for the $\pi$-conjugated systems \cite{cappelluti2012, szczesniak2025}, non-trivially capturing the quantum mechanical decay of orbital overlap with distance, which should be especially relevant in small, flexible molecules where electronic interactions are highly sensitive to bond-length fluctuations.

Interestingly, the introduced linear electron-phonon coupling can be additionally linked to the somewhat sibling reorganization energy \cite{atahan2018, hsu2020, hsu2022}, familiar in the {\it swipe card} model of olfaction \cite{brookes2007, brookes2012} and rooted in the Marcus theory \cite{marcus1956}. For this purpose, it is convenient to recall the potential energy function for the quantum oscillator $V(u)=1/2 M \omega^2 u^2$, where $M$ is the effective mass and $\omega$ stands for the characteristic frequency. Assuming that the final electronic state of the odorant shifts the potential minimum due to the linear electron-phonon coupling term, the reorganization term can be simply given as $\lambda=\alpha^2/2M\omega^2$. Here, it is assumed after Brookes \cite{brookes2007} that $M=8.25$ amu and $\hbar\omega=0.2$ eV. These can be considered representative values of the odorant molecules that also allow comparison with other estimates available in the literature. In this manner, the direct relation between the electronic coupling and two scales related to the vibrations of the odorant is established.

\begin{figure}[ht!]
\includegraphics[width=\columnwidth]{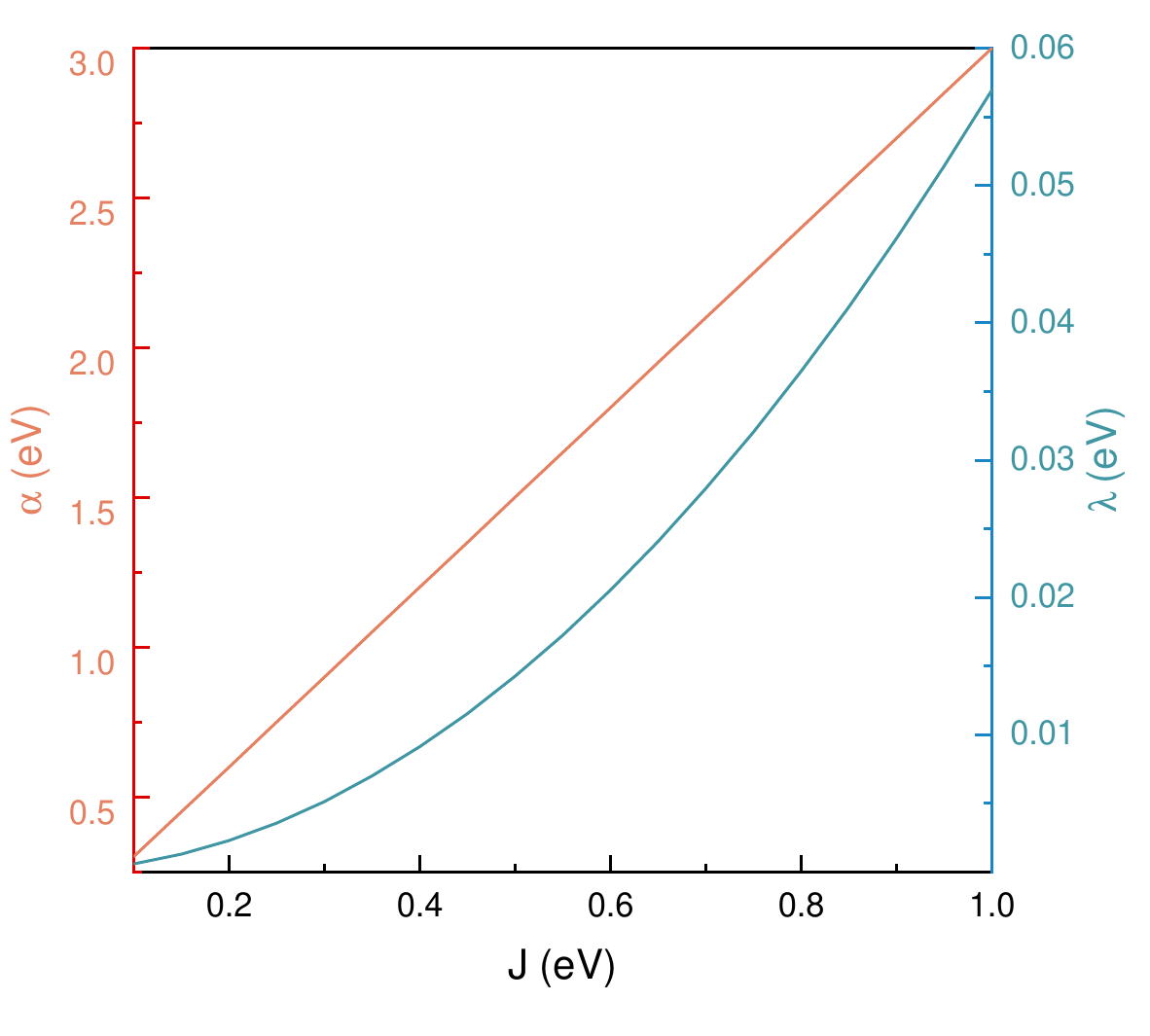}
\caption{The magnitude of the linear electron-phonon interaction (orange) and reorganization energy (turquoise) as a function of the electronic coupling (black).}
\label{fig03}
\end{figure}

In Fig. \ref{fig03}, the behavior of the linear electron-phonon interaction and reorganization energy as a function of the electronic coupling is presented. Note that only $J<1$ eV is considered, in agreement with the earlier condition for strongly localized HOMO and LUMO. In general, both discussed terms are clearly increasing along with the increase in the electronic coupling, although they do so in qualitatively distinct ways. The electron-phonon coupling exhibits linear strengthening, which is in line with the earlier observations made for carbon systems \cite{cappelluti2012, szczesniak2025} and reflects the exponential dependence of orbital overlap on interatomic distance. More precisely, since $J(u) \propto e^{-\eta u}$, its gradient with respect to displacement naturally scales with its own magnitude, yielding $\alpha \propto J$. This implies that stronger electronic tunneling leads directly to a proportionally enhanced coupling to nuclear motion. On the other hand, the reorganization energy shows a nonlinear (quadratic) increase and indicates that strong tunneling not only enhances the coupling to vibrations but also amplifies the structural response, making the system increasingly vibronically active. However, of particular importance to the presented discussion are the quantitative, not qualitative, features of the presented results. Specifically, it can be observed that for $J \sim$ 0.7 eV, the reorganization energy is equal to 0.03 eV, in agreement with the value suggested for odorant molecules by Brookes \cite{brookes2007}. In what follows, this reinforces the above suggestion that tunneling across odorant molecules likely falls into the deep off-resonant limit, supporting the argumentation behind the {\it swipe card} model of olfaction.

In summary, the general analysis presented here offers novel insights into the regime of quantum tunneling across odorant molecules. It establishes fundamental limits for such processes based on the intrinsic energy scales that govern these molecular systems, thereby shedding light on the mechanism of olfaction. In particular, in the context of the {\it swipe card} model of olfaction, presented results appear consistent with this framework. Specifically, they provide partial support for the idea that quantum tunneling may contribute to the sense of smell. While tunneling alone may not fully explain olfactory recognition, it should be considered as one of several contributing factors. To further investigate this direction, the study provides useful theoretical relations and identifies measurable observables that could be tested in future experiments. Moreover, the same framework may serve as a foundation for large-scale screening of odorant-receptor interactions using realistic molecular structures and advanced computational techniques. In particular, machine learning approaches might be valuable in validating and expanding upon the arguments presented in this study.

\begin{acknowledgements}
S.K. would like to acknowledge funding from the U.S. Department of Energy (DOE) Office of Basic Energy Sciences Award No. DE-SC0019215.
\end{acknowledgements}

\bibliography{bibliography}

\end{document}